# Freeform imaging system design with multiple reflection surfaces

Yunfeng Nie,[1] David R. Shafer,[2] Heidi Ottevaere,[1] Hugo Thienpont,[1] and Fabian Duerr[1]

[1]*Brussels Photonics, Department of Applied Physics and Photonics, Vrije Universiteit Brussel Pleinlaan 2, 1050 Brussels, Belgium*

[2]*David Shafer Optical Design, 56 Drake Lane, Fairfield, CT. 06824 USA*

**Abstract:** Reflective imaging systems form an important part of photonic devices such as spectrometers, telescopes, augmented and virtual reality headsets or lithography platforms. Reflective optics provide unparalleled spectral performance and can be used to reduce overall volume and weight. So far, most reflective designs have focused on two or three reflections, while four-reflection freeform designs can deliver a higher light throughput (faster F-number) as well as a larger field-of-view (FOV). However, advanced optical design strategies for four-reflection freeform systems have been rarely reported in literature. This is due to the increased complexity in solution space but also the fact that additional mirrors hinder a cost-effective realization (manufacture, alignment, etc.).

Recently, we have proposed a novel design method to directly calculate the freeform surface coefficients while merely knowing the mirror positions and tilts. Consequently, this method allows laymen with basic optical design knowledge to calculate 'first time right' freeform imaging systems in a matter of minutes. This contrasts with most common freeform design processes, which requires considerable experience, intuition or guesswork. Firstly, we demonstrate the effectiveness of the proposed method for a four-mirror high-throughput telescope with 250mm-focal-length, F/2.5 and a wide rectangular FOV of 8.5º × 25.5º. In a subsequent step, we propose an effective three-mirror but four-reflection imaging system, which consists of two freeform mirrors and one double-reflection spherical mirror. Compared with common three-mirror and three-reflection imagers, our novel multi-reflection system shows unprecedented possibilities for an economic implementation while drastically reducing the overall volume.

## Introduction

Imaging systems play a vital role in many photonics applications in consumer electronics[1], lithography[2-3], aerospace[4-5], safety and security[6], life science[7] or fiber-optics[8]. The development of these systems often starts with optical design. For decades, optical designers require considerable experience with a strong background in aberration theory and to be ideally familiar with a wide range of classic lens designs to perform this task. The rapid development in computing power and optimization algorithms have been paving the way towards more automated optical design approaches[9]. Present-day computers are used for raytracing and available optimization algorithms such as Damped Least Squares (DLS) or Orthogonal Descent (OD) are used to improve initial optical designs in an iterative process[10]. Consequently, such modern design work is still largely driven by time-consuming and often tedious trial-and-error work of raytracing-based optimization iterations to achieve a 'feasible' design. Thus, most optical designers can complete a sophisticated lens design task in reasonable time if they have access to a good 'initial design' to start with.

However, this routine work of 'initial design - iterative optimization' has been challenged by freeform optical systems due to the many more required variables to describe such surfaces and the limited availability of adequate initial designs. Growing interest in freeform optics is fostered by the rapid advances in precise fabrication techniques and replica of optical components with almost any shape[11-13]. Including freeform components in optical systems provides numerous opportunities for unprecedented performance and lightweight, compact packaging, which has already been proven in many applications such as spectrometers[14], Augmented/Virtual/Mixed Reality (AR/VR/MR)[1] and biomedical imaging[15]. To maximize the advantages of using freeform surfaces, there are typically four design strategies: **paraxial designs followed by further optimization**[1,16-17], **automated designs using advanced optimization**[18], **Nodal Aberration Theory based designs**[19] and **direct designs where certain aberrations vanish**[20-21].

The first and common design strategy relies on basic starting points which usually results from paraxial calculation, and subsequently using commercial optical design software (e.g. Zemax or Code V) for iterative optimization. The software can target different features of freeform optics design, for example, applying orthogonal surface representations to ease manufacturability or improve the optimization[22-23]. However, practical issues such as deviations from best fitting spheres (BFS), frequent occurrences of obscuration and overall volume are usually ignored by available merit functions. It is important to note that the paraxial assumptions in most optical design programs can lead to unreliable calculations of crucial metrics such as the effective focal length (EFFL) or distortion for freeform systems. The second strategy focuses on automated design by using advanced optimization methods with reduced required input by the optical designer. It usually relies on a point-by-point construction process to minimize a user-defined merit function in order to generate high-performance freeform imaging systems[18]. In contrast, Nodal Aberration Theory (NAT) provides an insightful and systematic prediction and visualization of the contributions of Zernike surface terms to the full-field aberrations[24-25]. As such, it can guide the experienced designer towards a successful design provided an in-depth understanding of the underlying theory. The fourth design strategy relies on solving geometrical or differential equations describing the freeform optical system under study to achieve a well-performing initial design[20,21,26]. Although these methods show clear merits, so far, they lack a straightforward path to increase the number of optical surfaces that can be calculated.

When it comes to four-reflection freeform imaging designs, the above-mentioned methods will typically reach a local minimum, but another designer might hardly reproduce the design without knowing the exact multi-step 'recipe'. Furthermore, the analytic solutions for unobscured four-mirror freeform systems has been considered as too algebraically complex[27], thus only few attempts have been made to directly construct four-mirror spherical/conic designs[28-31], eventually further optimized but limited to a relatively slow F-number (F#) and/or a moderate FOV[32-33].

In this paper, **we have extended our so-called 'first time right' direct design method** [34] for fast, wide-field four-reflection freeform systems, which allows to systematically expand the range of aberrations that need to be controlled or cancelled. By contrast, this analytic solution provides an exact set of solved parameters for each design, thus **one starting geometry** will always result in **the same deterministic result** with most aberrations corrected, leveraging the superiority of using freeform surfaces in high-performance imaging systems. The method allows to match user-defined conditions such as minimal packaging and minimal image blurring for individual aberrations orders by calculating the corresponding surface coefficients from scratch. Consequently, the method can prove beneficial for "junior" and "senior" optical designers alike and requires only basic optical design knowledge.

In a first example, we demonstrate the effectiveness of the proposed method for a 'classic' four-freeform-mirror telescope design with a wide field of view. In a subsequent step, we propose and investigate a class of three-mirror systems that can be equally characterized as four-reflection imaging systems. Such systems consist of three mirrors where one of the mirrors is seen twice with a double reflection. Compared with common three-mirror three-reflection imagers, the use of a double reflection could keep an excellent imaging performance in a more compact volume while reducing the manufacturing and alignment complexity.

## Mathematical description

*System coordinates and geometry.* Traditional refractive optical systems are usually axial symmetric (all components are perpendicular to the axis); thus, one axial length is typically enough to determine a lens position. Unobscured reflective imaging systems have multiple mirrors, where each is defined in a three-dimensional (3D) space with distances and tilt angles. While the definitions and procedures are generic to any reflective imaging system, we here show the example of a four-mirror system in Fig. 1. The system consists of four sequential mirrors M1, M2, M3, M4 and the image plane, where rays travel from the object space (left) to the image space. The surface positions are defined in a global coordinate XYZ, while each surface geometry is described by its local coordinate **uvs**. Like any imaging system, the design has an entrance pupil surface for sampling rays, which theoretically can be any position in the system and the current drawing is for illustrative purpose only. The origin of the global coordinate is put on the vertices of M1, while the Z-axis vector is following the local Z vector of the entrance pupil plane. Any freeform surface is spatially determined by its global vertices $(X_i, Y_i, Z_i)$ ($X_i$ can be ignored due to YZ-plane symmetry). The surface sag function $f_i$ is based on a local coordinate $(u_i, v_i)$ with its origin on the vertices, and the local coordinate axes are respectively the line orthogonal to YZ plane (u), the tangent line (v) and the normal line (s).

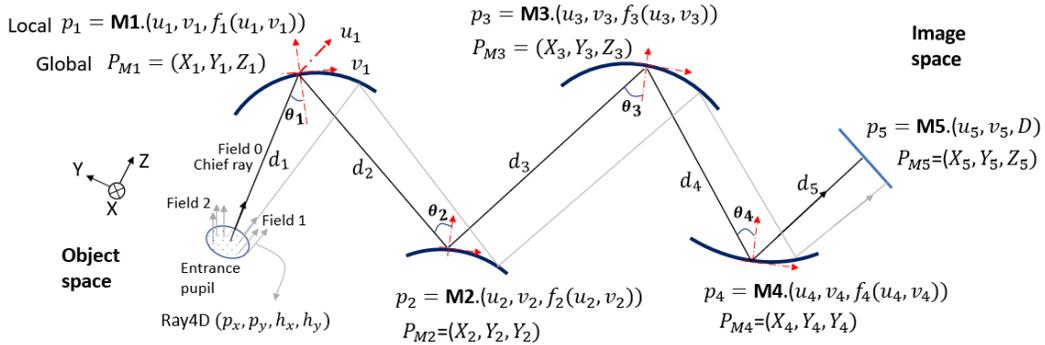

Fig. 1. Sketch of an unobscured four-mirror imaging system: with global coordinates **XYZ** to locate each surface, and local coordinates **uvs** to describe surface geometry.

*Entrance pupil sampling, fields and 4D ray.* Each ray can be represented by one point on the object plane and another point on the entrance pupil. On the entrance pupil, we sample the circular area by a certain pattern, e.g. using Gaussian quadrature with a given number of rings and arms[35]. Thus, each point is denoted locally by a normalized 2D-vector $(p_x, p_y)$. On the object plane, one object point emitting a bundle of rays is one field, described by a normalized 2D position vector $(h_x, h_y)$ if the object is from finite distance, otherwise described by a normalized 2D angular vector $(h_x, h_y)$. The proposed method is applicable for both situations, while this text will focus only on objects at infinity if not specified otherwise. Any ray entering the imaging system is solely determined by these two planes, forming a 4D ray $(p_x, p_y, h_x, h_y)$.

***Mapping functions and surface functions***. We now consider a sequence of N reflective optical surfaces $f_i$ (i=1…N) aligned along a principal ray path. The mapping functions $u_i, v_i$ describe the trajectories of the 4D rays by mapping them from the 1$^{st}$ surface to image plane (denoted as N+1), and represented by power series

$$u_i(p_x, p_y, h_x, h_y) = \sum_{j=0}^{J} \sum_{k=0}^{K} \sum_{l=0}^{L} \sum_{m=0}^{M} \bar{u}_{i,jklm} \, p_x^j p_y^k h_x^l h_y^m, \quad i = 1 \ldots N+1 \quad (1)$$

$$v_i(p_x, p_y, h_x, h_y) = \sum_{j=0}^{J} \sum_{k=0}^{K} \sum_{l=0}^{L} \sum_{m=0}^{M} \bar{v}_{i,jklm} \, p_x^j p_y^k h_x^l h_y^m, \quad i = 1 \ldots N+1 \quad (2)$$

To describe the surface functions, we use a power series representation

$$f_i(x,y) = \sum_{s=0}^{S} \sum_{t=0}^{T} \bar{f}_{i,st} x^s y^t, \quad i = 1 \ldots N \quad (3a)$$

Where, j, k, l and m denote the orders of pupil and field coordinate, and $\bar{u}_{i,jklm}, \bar{v}_{i,jklm}, \bar{f}_{i,st}$ are unknown coefficients. Once the $\bar{f}_{i,st}$ are solved, the freeform surfaces are also fully determined. With respect to manufacturing, we can separate the spherical part and the higher order terms to better quantify the freeform departure. As the defined expressions of power series are quite close to XY polynomials, we here use a freeform surface definition, which is mathematically represented by a spherical basis and a freeform departure

$$f_i(x,y) = \frac{c_i(x^2+y^2)}{1+\sqrt{1-c_i^2(x^2+y^2)}} + \sum_{m=0}^{M} \sum_{n=0}^{N} A_{i,mn} x^m y^n, \quad i = 1 \ldots N \quad (3b)$$

where, $c_i$ is the paraxial curvature of the surface and $A_{i,mn}$ are coefficients of the polynomial. Given the yz-plane symmetry, only even x-terms are required. Once the coefficients $\bar{f}_{i,st}$ are known, the surface can be converted, e.g. to Zernike polynomials, Forbes Q polynomials and alike.

***Ray aberration expansions***. The purpose of an optical system is to propagate arbitrary rays to its ideal image points $(u_{ideal}, v_{ideal})$, that is

$$\begin{cases} u_{ideal} = F_x \cdot \tan(h_x), \; v_{ideal} = F_y \cdot \tan(h_y), \; \text{for infinite object} \\ u_{ideal} = M_x \cdot h_x, \quad\quad\; v_{ideal} = M_y \cdot h_y, \; \text{for finite object} \end{cases} \quad (4)$$

$F_x$ and $F_y$ are the focal lengths in X- and Y-direction, $M_x$ and $M_y$ are the magnifications in X- and Y-direction, respectively. Thus, the deviation of an arbitrary ray from its ideal image point is $(u_{N+1} - u_{ideal}, v_{N+1} - v_{ideal})$, which we express as a power series $\epsilon = (\epsilon_x, \epsilon_y)$

$$\epsilon_x(p_x, p_y, h_x, h_y) = \sum_{j=0}^{J} \sum_{k=0}^{K} \sum_{l=0}^{L} \sum_{m=0}^{M} \bar{\epsilon}_{x,jklm} \, p_x^j p_y^k h_x^l h_y^m \quad (5)$$

$$\epsilon_y(p_x, p_y, h_x, h_y) = \sum_{j=0}^{J} \sum_{k=0}^{K} \sum_{l=0}^{L} \sum_{m=0}^{M} \bar{\epsilon}_{y,jklm} \, p_x^j p_y^k h_x^l h_y^m \quad (6)$$

to represent the transverse ray aberrations in X- and Y-direction, where $n = j + k + l + m$ denotes the aberration order n.

## Evaluation and optimization metrics

In the following, we introduce relevant metrics that are essential for a thorough system evaluation and optimization. Here, we assume that all relevant mapping coefficients $\bar{u}_{i,jklm}, \bar{v}_{i,jklm}$, surface coefficients $\bar{f}_{i,st}$ as well as aberration coefficients $\bar{\epsilon}_{x,jklm}, \bar{\epsilon}_{y,jklm}$ have been already calculated successfully.

***System layout and volume.*** The freeform surface sag in local coordinates is constructed following Eq. (3a). To visualize the 2D layout of the entire optical system, a grid of points on the calculated freeform surfaces is sampled to form 3D point clouds, which are converted to global coordinates. We extract the point clouds over the x=0 plane and plot it to generate the 2D layout.

In order to calculate the packaging volume, we sample the 4D rays over the full FOV and full entrance pupil to get the global coordinates of boundary points. Then, the maximum and minimum values of all surfaces are determined as $X_{min}, X_{max}, Y_{min}, Y_{max}$ and $Z_{min}, Z_{max}$, respectively. The smallest enclosed volume is defined as the volume of the minimum bounding box that can put the erect system inside (H × W × L):

$$V_{3d} = (Z_{max} - Z_{min}) \times (Y_{max} - Y_{min}) \times (X_{max} - X_{min}) \tag{7}$$

***Freeform departures from best fitting sphere.*** Freeform departures from besting fitting sphere are one possible indication of the difficulty and cost in manufacturing a freeform mirror. As the surface function is known, we can readily obtain the best fitting sphere (BFS) radius $c_i$ by using existing algorithms, for example, the Fast Sphere Fit by Sumith YD[40]. The freeform departure contour is calculated by

$$\Delta f_i(x,y) = \sum_{s=0}^{S}\sum_{t=0}^{T} \bar{f}_{i,st} x^s y^t - \frac{c_i(x^2+y^2)}{1+\sqrt{1-c_i^2(x^2+y^2)}}, i = 1 \dots N \tag{8}$$

***Root-Mean-Square spot size.*** Besides the system feasibility and manufacturability, the image quality needs to be quantified. The root-mean-square (RMS) spot size and wavefront error are both comprehensive criteria that consider the superposition of primary and high-order aberrations. In this work, we mainly use RMS spot sizes as the image sharpness criterion. The RMS spot size is assessed over a full field of view based on a specified sampling strategy, such as cartesian sampling and polar sampling[35]. We use polar sampling with arms $n_a$ and rings $n_r$ defined to determine the ray number. From Eqs. (5)-(6), we obtain the radial aberrations for an arbitrary ray

$$\epsilon(H) = \sqrt{\epsilon_x^2(p_x, p_y, H) + \epsilon_y^2(p_x, p_y, H)}, \text{ with } p_x^2 + p_y^2 \leq 1 \tag{9}$$

Where H represents one normalized field $(h_x, h_y)$ according to FOV definition

$$\begin{cases} H: -1 \leq h_x \leq 1, -1 \leq h_y \leq 1, \text{in rectangular FOV} \\ H: h_x^2 + h_y^2 \leq 1, \text{in circular FOV} \end{cases} \tag{10}$$

We then perform the RMS calculation by evaluating rays in $p_x$ and $p_y$ for one specified field H

$$\text{RMS}(H) = \sqrt{\frac{1}{n_a n_r}\left(\sum_{i=1}^{n_a}\sum_{j=1}^{n_r}\epsilon_x^2(p_{x,ij}, p_{y,ij}, H) + \sum_{i=1}^{n_a}\sum_{j=1}^{n_r}\epsilon_y^2(p_{x,ij}, p_{y,ij}, H)\right)} \tag{11}$$

The RMS spot size metric is calculated as the average RMS value over the selected fields. Weighting factors can be furthermore added to improve the image quality for specified fields.

***Distortion.*** Distortion is an aberration that does not affect the image sharpness but causes image deformation. Current digital processing techniques provide good corrections, but in optical design there are typically minimum requirements. Following Eqs. (4)-(6), the distortion is calculated by the deviation of the chief ray $(p_x = 0, p_y = 0)$ from the ideal image in percentage

$$\text{DIST}(H) = \frac{\sqrt{\epsilon_x^2(0,0,H) + \epsilon_y^2(0,0,H)}}{\sqrt{u_{ideal}^2(H) + v_{ideal}^2(H)}} \times 100\% \tag{12}$$

As the distortion is a function of the field, the distortion metric value is defined as the maximum distortion over the selected fields.

***Obscuration.*** Another problem that occurs frequently during the design of off-axis mirror systems is obscuration. Without intervention, an off-axis design is easily lost into an impractical obscured system which can be a substantial bottleneck to overcome. We have defined an obscuration metric as a penalty term for the optimization merit function to address this problem. The metric is universal to multi-mirror systems, e.g. three- or four-mirror systems.

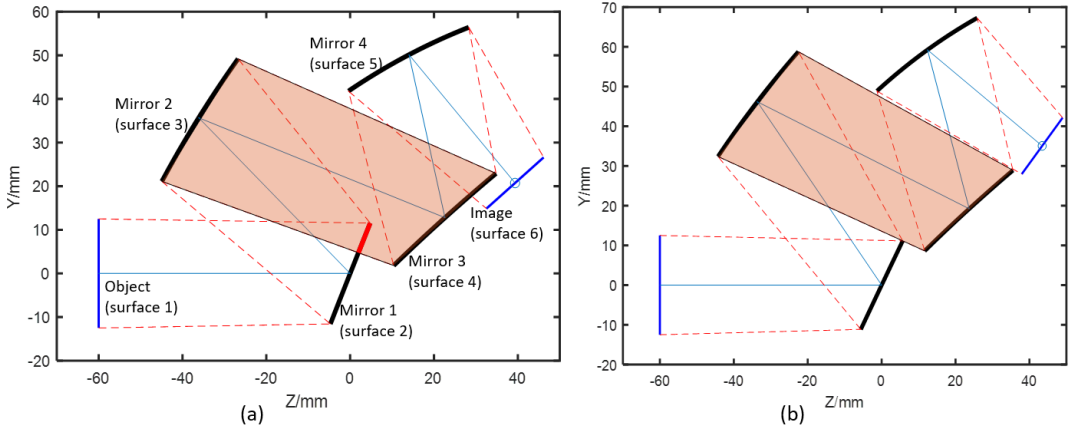

Fig. 2. Use obscuration metric to ensure zero-obscuration: (a) an initially partially obscured mirror system and (b) an unobscured system after optimization by using the obscuration penalty term.

From the 'first time right' calculations, we can generate the system layout as shown in Fig. 2(a), throughout surfaces 1 to N+1. Only the mirrors and the image plane are real surfaces while the others are virtual lines. One region is defined as two adjacent real surfaces (in solid lines) as well as the virtual in-between boundaries (in dotted lines). One overlap segment is detected once a third surface is intersected with the region, highlighted in red bold lines. Traversing the real surfaces in the region, the overall overlap is the sum of all the intersected segments:

$$\text{overlap(i)} = \begin{cases} \sum_k \text{length}(k), & k \text{ is the number of intersected segments} \\ 0, & \text{no intersected segments} \end{cases} \quad (13)$$

By traversing all the regions in the system layout, the obscuration metric is calculated as the sum over overlap(i):

$$\text{obscuration} = \sum_{i=1}^{N-1} \text{overlap}(i) \quad (14)$$

Figure 2 shows the effective automatic obscuration removal: the 2D layout after adding the obscuration metric into optimization process (right) compared with no obscuration metric (left).

## 'First time right' optical design procedure

The procedure of the 'first time right' calculations is summarized, whereas more details can be found in our recent article[34]. First, the system geometrical layout and specifications such as EFFL, entrance pupil diameter (ENPD) and position, maximum FOV, distortion and volume constraints are predefined. An arbitrary ray path from object to image plane thus consists of (N+1) segments with corresponding distances $d_1 \ldots d_{N+1}$. By applying Fermat's principle to pairs of consecutive optical path segments $D_{i,x} = \partial_{u_i}(d_i + d_{i+1}) = 0$, $D_{i,y} = \partial_{v_i}(d_i + d_{i+1}) = 0$; $i = 1 \ldots N$, we can derive 2N partial differential equations that can be solved using a power series method.

$$\lim_{p_x \to 0} \lim_{p_y \to 0} \lim_{h_x \to 0} \lim_{h_y \to 0} \frac{\partial^j}{\partial p_x^j} \frac{\partial^k}{\partial p_y^k} \frac{\partial^l}{\partial h_x^l} \frac{\partial^m}{\partial h_y^m} D_{i,x} = 0, \ i = 1 \ldots N \quad (15)$$

$$\lim_{p_x \to 0} \lim_{p_y \to 0} \lim_{h_x \to 0} \lim_{h_y \to 0} \frac{\partial^j}{\partial p_x^j} \frac{\partial^k}{\partial p_y^k} \frac{\partial^l}{\partial h_x^l} \frac{\partial^m}{\partial h_y^m} D_{i,y} = 0, \ i = 1 \ldots N \quad (16)$$

With the underlying differential equations established and all system specifications introduced, there are two calculation steps that need to be performed:

(1) Solve the non-linear first order case for j+k+l+m=1 given Eqs. (15) and (16), either by using a user-defined non-linear solver or by making use of alternative matrix optics tools [31, 36]. This calculation results in a set of nonlinear equations for second order surface coefficients and first order mapping and aberration coefficients. Extra conditions for the second order surface coefficients can be imposed if desired.

(2) For each higher order o=j+k+l+m=2,3…, we can obtain the exact linear relationship among the surface, mapping and aberration coefficients for all corresponding indices (i,j,k,l,m) once the previous order has been solved. Each linear equation set of order o relates to surface coefficients $\bar{f}_{i,st}$ with s+t=o+1 while the mapping coefficients $\bar{u}_{i,jklm}, \bar{v}_{i,jklm}$ and aberration coefficients $\bar{\epsilon}_{x,jklm}, \bar{\epsilon}_{y,jklm}$ have j+k+l+m=o. Then we sort them in ascending order by setting unwanted aberrations to zero or by minimizing a combination thereof as required by the targeted specifications of the imaging freeform system. For each order, we apply the elimination method for solving linear systems to eliminate the unknown mapping coefficients and to obtain a reduced linear system that expresses the aberration coefficients as linear functions of the unknown surface coefficients of that order. We then define a basic set of weighting factors $WF_{jklm} = (ENPD/2)^{j+k}(FOV/\sqrt{2})^{l+m}$ that is multiplied with the reduced linear equations of same index (j,k,l,m). The weighted least-squares solution for the reduced linear system then takes both the maximum entrance pupil diameter and largest diagonal field diameter into account to simultaneously minimize all properly weighted aberrations for each order, and to calculate the corresponding surface and aberration coefficients. The calculated coefficients are now substituted into the original linear system to calculate the remaining unknown mapping coefficients of that order.

In addition, the used power series solution method enables a function-based analytic ray tracing by accurately calculating higher-order mapping and aberration coefficients in ascending order and for all combinations of j+k+l+m=1,2,3… up to a desired order that is typically higher than that of the preceding surface coefficient calculations. Based on these 'first time right' calculations, a merit function based on targeted metrics can be generated. The subsequent rapid optimization process continues to leverage the performance by varying the mirrors' positions and tilts within certain ranges while minimizing two types of metrics: the image quality metrics (full-field RMS spot size and distortion) and the feasibility metrics (enclosed volume, obscuration and freeform departures from BFS). The overall design process flow chart is summarized in Fig. 3.

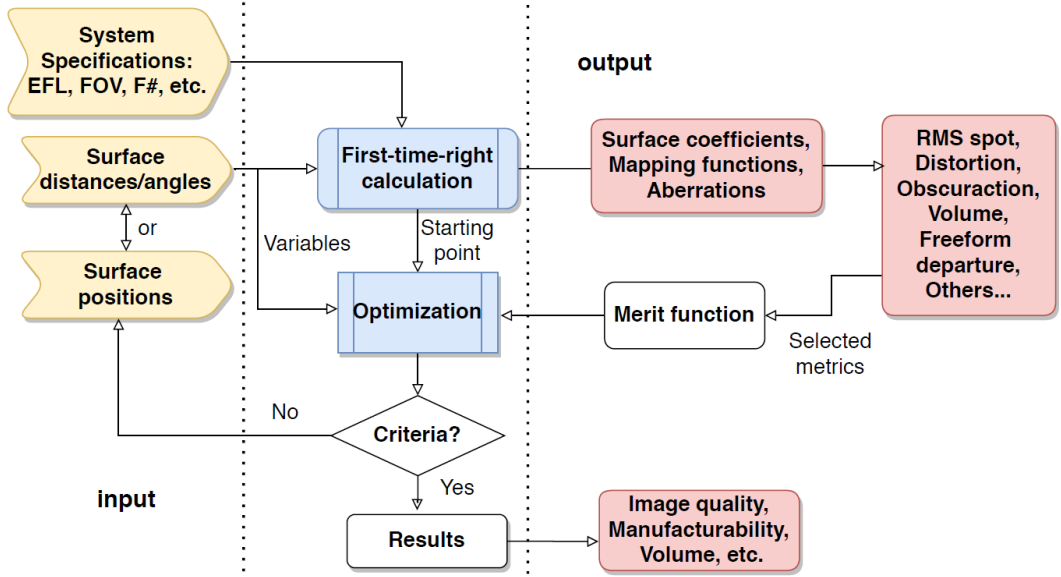

Fig. 3. Flowchart of the overall design process, which highlights the minimal required input by the user, and the wide range of directly calculated coefficients and metrics by the proposed method.

The weighting can be adjusted to first remove any overlap. Afterwards, the optimizer will further improve the performance through metrics such as the RMS spot size and the freeform departures. Finally, the obtained 'first time right' solution can be further fine-tuned when introduced in a classical ray-tracing software and applying the embedded advanced optimization algorithms.

## Design example 1: a fast four-mirror design with wide FOV

Mirror-based imaging systems are well-known for their application in telescopes where a long EFFL and large entrance pupil diameter (or high resolution) are the two most common specifications. Typically, the FOV is quite limited in such systems, e.g. most three-mirror-anastigmat (TMA) systems have an FOV of less than 10 degrees[18, 40]. As the FOV increases, astigmatism and field curvature dominate which are quite difficult to correct. Another challenge for large FOV, compact mirror systems is that a fast f-number is typically required to ensure high resolution while the optimization process will easily result in not feasible overlaps among mirrors. Adding the obscuration removal metric into the optimization process tends to enlarge the volume dramatically, thus a volume constraint is also needed; however, these two requirements can be in conflict that needs to be dealt with prior to the optimization process. To demonstrate the capability of the proposed method in achieving both high image performance and feasibility, we identified a state-of-the-art design with challenging specifications, listed in Table 1 (see reference[41]).

Table 1. Given specifications of the first design example

| Specifications | Parameters | Specifications | Parameters |
|---|---|---|---|
| Focal Length (mm) | 250 | Wavelengths (μm) | 0.5-5 |
| Field of view (º) | 25.5 x 8.5 | RMS spot diameter (μm) | < 18 |
| F-number | 2.5 | Distortion (%) | < 4 |
| Enclosed volume (mm) | 281x544x468 (70L) | Telecentricity(º) | < 2 |

Our method shows clear advantages for packaging requirements, since we only need to choose the five positions of initial planes to fulfill the physical size requirement in the first step. As a showcase, we define (0, 0), (-200, -115), (-180, 130), (-460, 20) and (-200, 245) as the global coordinates (x=0 for all vertices) of the four mirrors and the image plane respectively. Then we can determine the tilt angle for each surface to make the on-axis chief ray go through those points. The layout of five surfaces is determined in a global coordinate as shown in Fig. 4(a), which is a classic zigzag four-reflection system. The focal lengths for x and y dimensions are 250 mm with the stop placed at M3. With all system specifications established, we can initiate the calculation process as described in the previous section:

(1) We evaluate all $j + k + l + m = 1$ and $i = 1 \ldots 4$ for Eqs. (15) and (16) that result in a nonlinear system of 16 non-vanishing equations with 24 unknowns. Setting the four first order ray aberration coefficients $\bar{\epsilon}_{x,1,0,0,0}, \bar{\epsilon}_{x,0,0,1,0}, \bar{\epsilon}_{y,0,1,0,0}, \bar{\epsilon}_{y,0,0,0,1}$ to zero leaves 12 mapping coefficients $\bar{u}_{i,1,0,0,0}, \bar{u}_{i,0,0,1,0}, \bar{v}_{i,0,1,0,0}, \bar{v}_{i,0,0,0,1}$ ($i = 2,3,4$) and 8 second order surface coefficients $\bar{f}_{i,st}$ ($s + t = 2, i = 1 \ldots 4$) as unknowns. We further define four input variables $c_{i,xy}$ (default as 1) that link the 2$^{nd}$ order coefficients in x- and y-direction of the respective mirrors, that is $\bar{f}_{i,0,2} = c_{i,xy} \cdot \bar{f}_{i,2,0}$. With those definitions, the nonlinear equations are ready to be solved.

(2) We calculate the surface coefficients up to 6$^{th}$ order and the mapping and aberration coefficients up to 5$^{th}$ order, as outlined in the previous section. These calculations result in 52 surface coefficients, 375 mapping and 125 aberration coefficients. Note that several of the mapping and aberration coefficients are interdependent. The linear solution scheme of the mapping and aberration coefficients is further increased to 8$^{th}$ order (or higher if needed) to ensure a sufficiently accurate evaluation of the system performance.

The result of this initial calculation is shown in Fig. 4(b). Here, the obscuration has already been removed by manually adjusting the mirror positions, and the following optimization process maintains the zero-obscuration condition while the values of $c_{i,xy}$ ($i = 1 \ldots 4$) have been set to 1 initially. The corresponding RMS spot diagram of nine selected fields (three equal-spacing fields in X: -4.25º, 0º, 4.25º and +Y: 0º, 6.375º, 12.75º due to XZ-plane symmetry) is shown in Fig. 4(c). As we can see, the RMS spot radius of on-axis field is already as small as 13.23µm while the maximum field has an RMS spot radius of 306.05µm. The distortion is uncontrolled in the beginning with a value of 6.6%, and the maximum freeform PV departure is about 3343 µm. At this step, we have quickly reached a reasonable first layout with zero obscuration by manually adjusting the mirror positions while keeping an acceptable performance.

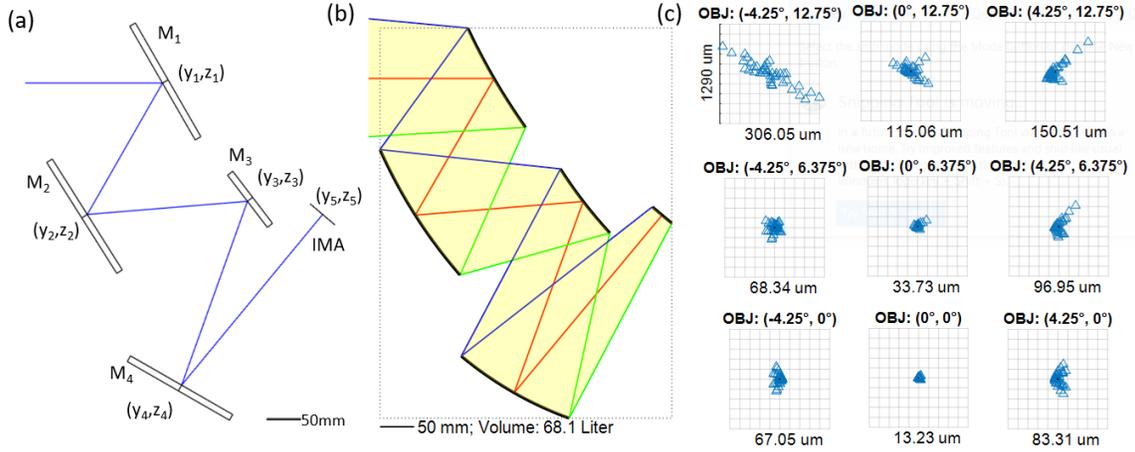

Fig. 4 The large FOV, fast four-mirror design example (a) global coordinates of initial planes (b) the layout after manually adjusting the mirror positions (c) the spot diagrams of nine selected fields and their RMS radii.

To further improve the imaging performance while keeping a feasible layout and moderate manufacturability, the metrics of total overlap, distortion, full-field RMS spot sizes, maximum freeform PV departure and smallest enclosed volume are added to the integrated merit function. The only variables are the positions of all surface vertices, varying within given ranges, e.g. ±10mm. The maximum distortion is set to be 4.3%. As distortion can be corrected by digital image processing techniques; this value could be loosened if otherwise the image quality must be compromised. The volume is bounded to be within 70L, and the maximum telecentricity angle should be lower than 2º for uniform illuminance. The allowable freeform PV departure for all four mirrors is set to be within reasonable manufacturing limits (500μm) described by Hentschel et al[43].

After a 500-iteration optimization which takes less than five minutes, both the image quality and the fabrication related metrics have been clearly improved. The merit function value decreased very quickly during the first 100 iterations, which takes only 75 seconds, as shown in Fig. 5(a). The layout in Fig. 5(b) is not far off the initial layout, while the vertices of five planes have been optimized to (0, 0), (-197.575, -121.548), (-177.513, 117.47), (-458.666, 17.491) and (-196.236, 258.281) mm respectively. The maximum RMS spot radius results in 27.56μm, about one-tenth of the value for the manually selected initial parameter set; and the smallest RMS spot radius is 0.92μm, already well below the Airy Radius of about 6 um at central wavelength 2μm (black circles in Fig. 5(c)). The freeform departures are shown in Fig. 5(d), with the four mirrors PV freeform deviations of 209 μm, 361 μm, 44 μm and 143 μm respectively, about ten times less than for the manual-selected initial design. The maximum distortion is 4.3%, note that the distortion can be corrected to below 4%, but the maximum freeform PV departure becomes 687μm compared to a much lower 381 μm if we keep a slightly looser distortion control.

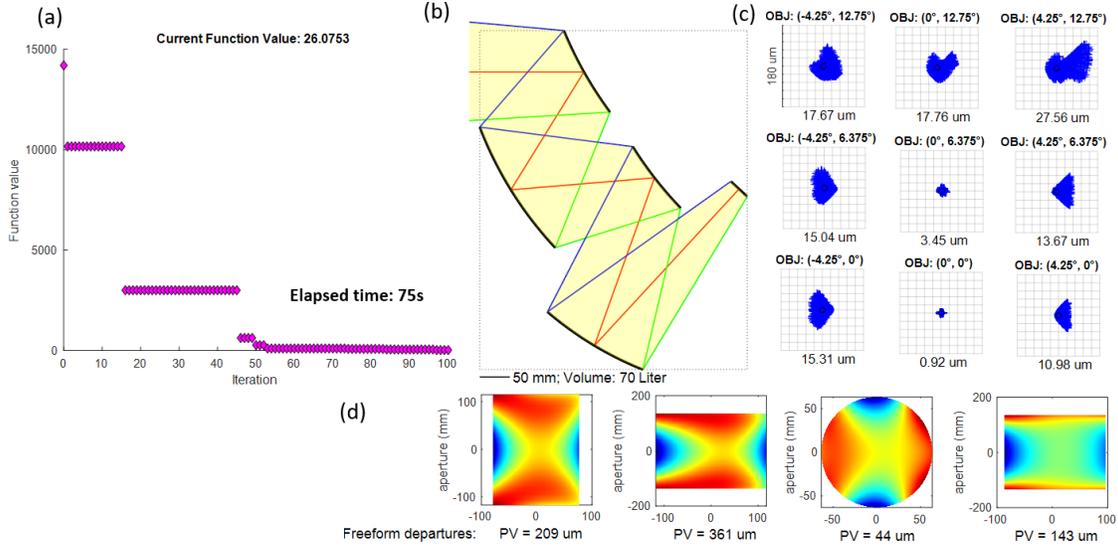

Figure 5. (a) The first 100-loop analytic optimization greatly improves the design. The evaluation after 500-loop analytic optimization: (b) the 2D layout (c) the spot diagrams of nine selected fields and their RMS radii and (d) the freeform departures of four mirrors from their BFS.

We then imported all positions, surface tilts and coefficients of the four mirrors into Zemax to verify the calculated system and further improve the image sharpness by using its DLS optimization. All the system specifications, the selected fields, the plane positions and tilts (four mirrors and the image plane) are kept the same. We constructed the merit function by using the optimization wizard tool and the PV Spot X+Y referring to centroid ray in 16 rings and 12 arms. The maximum distortion is constraint to stay below 4.3%. Only the surface coefficients are set as variables for the optimization to maintain the layout, but solely tweak the surface coefficients to further improve image sharpness. The optimization took about 6 minutes, and the RMS spot performance, full field wavefront error and the freeform departure from BFS are visualized in Fig. 6(a)-(c) respectively. The RMS spots are more balanced among all the fields, the maximum RMS spot is improved from 27.56μm to 4.89μm, while the best RMS spot is compromised from 0.92μm to 2.28μm. Note that the Airy Radius is 6.0μm, all fields are within diffraction limit. The full-field RMS wavefront error further describes the overall performance over 55 selected fields, close to a diffraction-limited design according to the Rayleigh Criterion[43].

The enclosed volume has slightly increased to 72L (no control in Zemax), and the freeform departure from BFS for each surface became slightly larger: the maximum PV departure has increased from 361μm to 457μm, which is still an excellent value in comparison to the reference design's 1053μm[41].

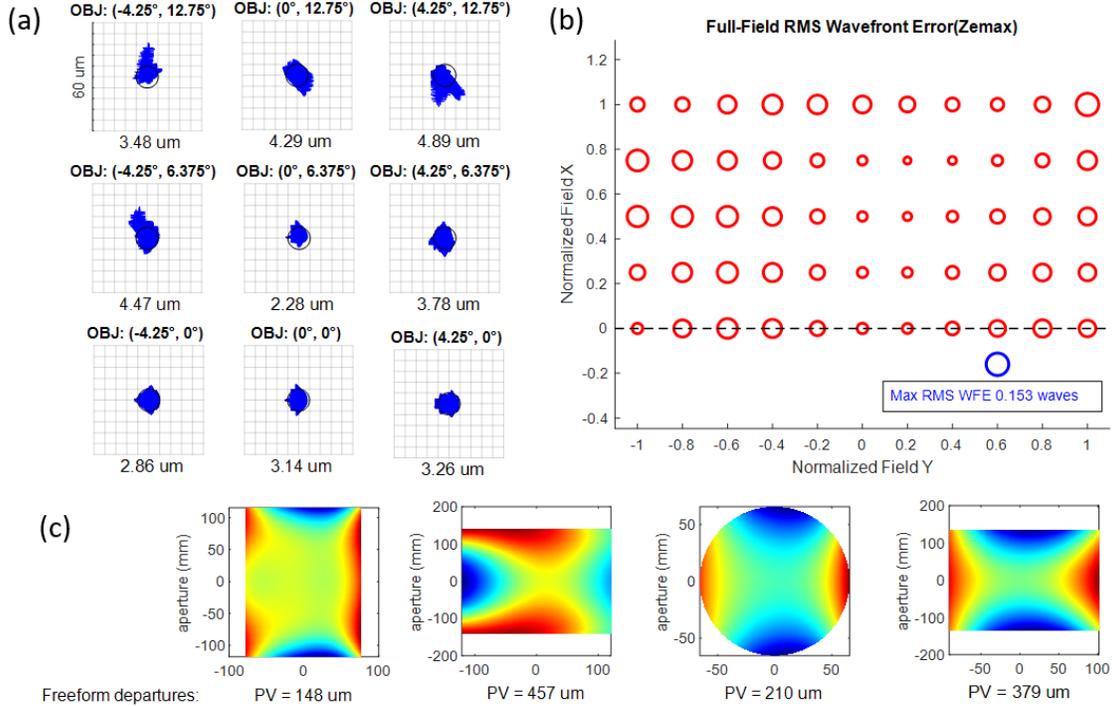

Figure 6. The system evaluation after Zemax optimization in terms of (a) the spot diagrams of nine selected fields and their RMS radii (b) the full-field RMS wavefront error and (c) the freeform departures of the four mirrors from their BFS.

## Design example 2: a two-freeform, three-mirror, four-reflection system

In practice, whether using a three-mirror or a four-mirror design is quite a substantial decision to make given the added extra complexity. As a potential and promising compromise, we have identified a new type of multi-reflection mirror systems, first proposed over forty years ago by Dave Shafer[44]. The idea is that at least one of the mirrors is passed twice in the ray path with a wanted overlap[45]. In this work, we revisited this brilliant idea for the first time to our knowledge using tailored freeform optical solutions. It is important to differentiate between our multi-reflection case and well-known Offner designs in literature where two individual mirrors share a joint substrate[46]. From a manufacturing and assembly point of view, a three-mirror imaging system with four or more reflections remains a three-mirror system. However, the proposed multi-reflection concept offers optical designers a largely extended design parameter space to explore and seek competitive solutions, e.g. to substitute the usage of curved image sensors.

Recently, **curved image sensors** have gained more attention for its unique advantages in compensating field curvature thus improving the image performance[47-49], but it can be a double-edged sword since they are still costly and not widely fabricated. In a competitive design, the usage of a spherical-curved sensor can achieve comparable performance[50]. Here, we demonstrate that the proposed method can also be applied to calculate such a new class of three-mirror, four-reflection systems, which can remove the necessity of using a curved sensor (toroid shape) compared to a state-of-the-art design while keeping comparable image quality[40]. The system specifications are listed in Table 2.

Table 2. Given specifications of the second design example

| Specifications | Parameters | Specifications | Parameters |
|---|---|---|---|
| Focal Length (mm) | 250 | Wavelength (μm) | 0.5 |
| Field of view (º) | 7.2 x 7.2 | RMS spot radius* (μm) | < 5 |
| F-number | 2.5 | Distortion (%) | < 1.33 |
| Enclosed volume (mm) | 226x406x514 (47L) | | |

* The result was achieved by using a toroidal sensor. If a flat sensor is used, the design has much larger RMS spot radii (typically >20 μm).

Following the proposed four-reflection 'first time right' method, we now force the second and fourth mirror to have one common surface radius and center to form a single double-reflection mirror. To better initialize the design without imposing constraints on the fourth reflection vertex, we introduced relative distances and angles as parameters to describe the design layout, which are defined as: $d_1 = 200$ mm, $t_1 = -0.36$, $d_2=250$ mm, $t_2=-1.116$, $t_3=-1.64$, $d_4=227$ mm, $d_3$ and $t_4$ are calculated from known parameters. Here, $d_i$ and $t_i$ are the distances between the mirrors and their respective tilt angles in radian as shown in Fig. 7(a). The angle is defined as the relative angle from +Y direction to the mirror tangent line at its vertices, which is negative if clockwise, and vice versa. It is important to emphasize that the 'first time right' calculation steps for four mirrors have been altered in order to address the double-reflection on one specified mirror:

(1) We evaluate the first order equations that result in a nonlinear system of 16 non-vanishing equations. Setting the four first order ray aberration coefficients to zero leaves 12 mapping coefficients and 4 second order surface coefficients as unknowns, that is $f_{i,2,0}$, $f_{i,0,2}$ ($i = 1,3$), whereas the base curvature $f_{2,0,2} = f_{2,2,0}$ at the joint mirror is one extra parameter. The nonlinear system is thereon ready to be solved.

(2) We calculate the surface coefficients up to 6$^{th}$ order for the surfaces and up to 8$^{th}$ order for the mapping and aberration coefficients while keeping the joint mirror strictly spherical.

The 2D layout after 100-iteration optimization of the distances and tilt angles is shown in Fig. 7(b), which takes 76 seconds, and has a volume of 23.9 L. This optimization allows the predefined surface positions to change within a certain range while keeping zero obscuration and its volume lower than 25 L. The spot diagram in Fig. 7(c) shows a well-corrected system, as the maximum RMS spot radius is 12.08 μm and the minimum RMS spot radius is 1.87 μm. The resulting design consists of one spherical mirror (M2, R=8260 mm) and two freeform mirrors (M1 and M3). The freeform departures of the two freeform mirrors from their best fitting spheres are shown in Fig. 7(d). The PV departures of 361μm and 316μm for M1 and M3 respectively are within the range of typical freeform manufacturability[42].

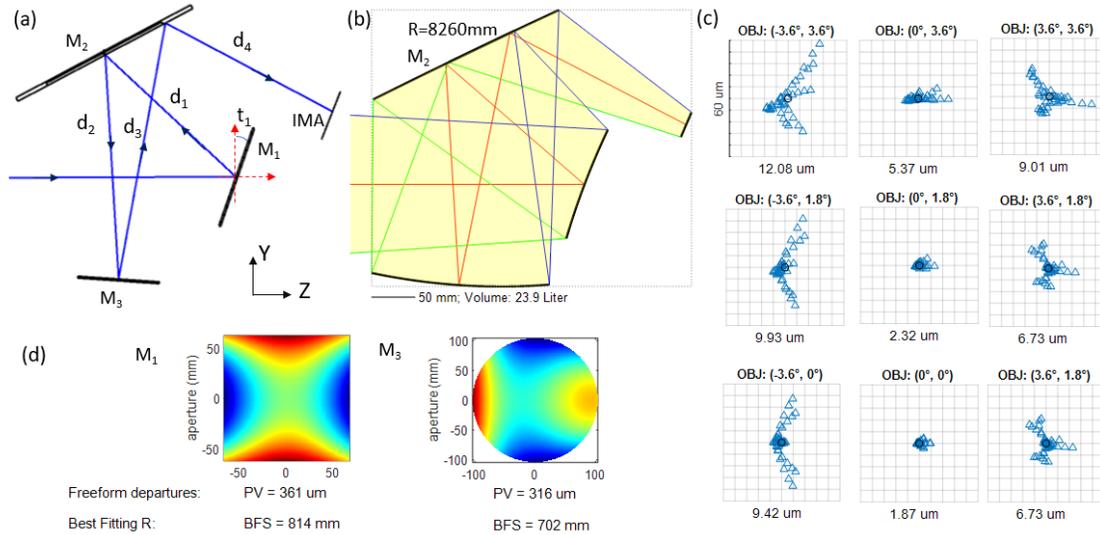

Fig. 7. The four-reflection, three-mirror, two-freeform mirror design using the proposed design method (a) The initial input parameters for the designer (b) the layout after 100-iteration optimization of the distances and tilt angles (c) the spot diagrams of nine selected fields and their RMS radii (d) freeform PV departures from their BFS for mirrors M1 and M3, note that M2 is a spherical surface with R = 8260 mm.

To verify the calculated design, we exported all three surfaces into Zemax, and a further optimization is performed to tweak the design, mainly to realize a more balanced RMS spot diagram. We used the default merit function: RMS spot X+Y referring to chief ray, pupil integration over 16 rings and 12 arms. DIMX is added to control distortion over different fields, as well as the maximum aspheric sag to control PV departures from BFS to not exceed 600 µm. The working f-number WFNO is set to be 2.5 and the entrance pupil 100 mm ensure an effective focal length of 250 mm. All surface positions are frozen to keep the zero-obscuration layout, only the coefficients of the two freeform surfaces plus the one spherical radius are set as variables. After about five-minute optimization, the merit function value stopped decreasing and the optimization converged. The final system is shown in Fig. 8(a), and its imaging performance in terms of RMS spots is shown in Fig. 8(b), where the maximum RMS spot radius is 4.69µm and minimum RMS spot radius is 2.65µm. Illustrated in Fig. 8(c), the freeform PV deviations of the two freeform mirrors decreased to 353µm and 228µm while the power on these mirrors increased. Note that the freeform departure contours are quite like those generated by the 'first time right' design in Fig. 8(c), underlining the excellent capability of the proposed method in finding an initial solution. Distortion seems to be difficult to be further corrected without compromising the image quality. As the location of stop is only feasible on M1 and M3, thus the configuration has lost certain symmetry over the stop, which causes a slightly worse distortion (2%) correction in comparison to the three-mirror system (1.33%). However, the still relative low distortion can be easily corrected by using digital processing, thus we consider it a more than reasonable compromise.

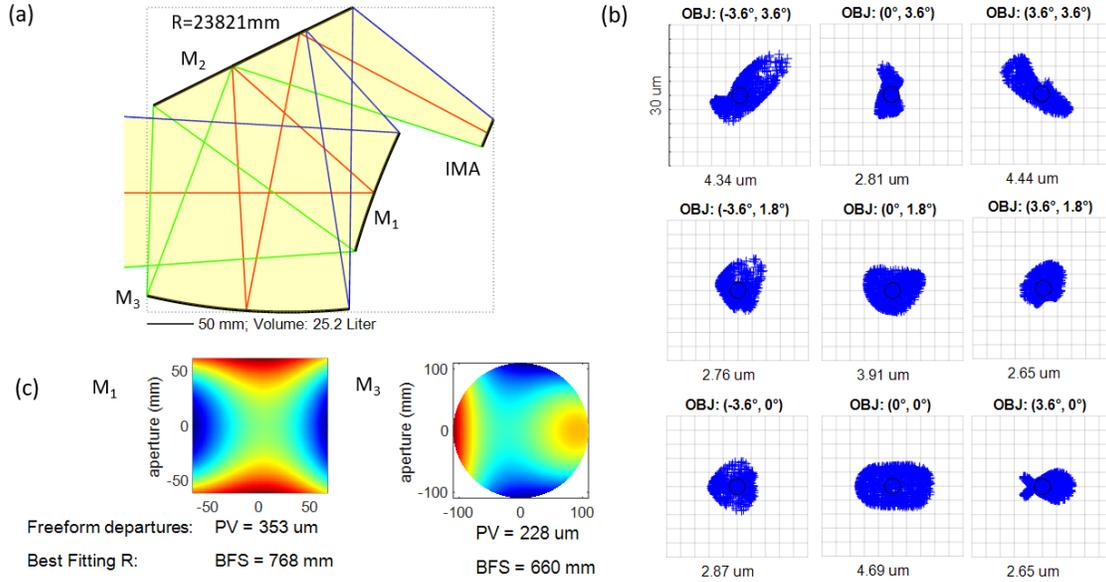

Fig. 8. Results for the four-reflection, three-mirror, two-freeform design after optimization in Zemax: (a) the 2D layout cross section (b) the spot diagrams of nine selected fields and their RMS radii (c) M1 and M3 freeform departures from their BFS; M2 remains as spherical surface with R = 23821 mm.

Compared with the reference designs[40], the volume is greatly improved by 50%. As for manufacturability, we use only two freeform mirrors in comparison to three freeform mirrors in the counterpart design, while the maximum freeform PV departure value is about 66% less, demonstrating a much higher feasibility from alignment and lower cost on the component fabrication. Besides, we use a flat image plane while the reference design has a toroid CMOS sensor to compensate for field curvature. As for image quality, the spot size indicates an approximately equal performance, and the distortion is slightly worse which is not significant and can be corrected by using digital processing. The complete comparison in terms of volume, RMS spot size, distortion, freeform PV departure and sensor shapes is summarized in Table 3. Since the focal length, FOV, F# and the wavelengths are the same, so these values are not listed.

Table 3. Final comparison of image quality and manufacturability

| Specifications | | Ref. design 1[40] | Ref. Design 2[40] | Our design |
|---|---|---|---|---|
| Surface type | | Legendre | Zernike | XY polynomial |
| Volume (liter) | | 49.1 | 47.2 | 25.2 |
| RMS spot radius (μm) | Max | 1.4 | 5.8 | 2.65 |
| | Min | 5.9 | 9.1 | 4.69 |
| | Mean | 3.65 | 7.45 | 3.5 |
| Distortion(max) | | 1.33% | 1.38% | 2% |
| Freeform PV departure from BFS (μm) | | M1:105.5<br>M2:1098.9<br>M3:354.9 | M1:128.1<br>M2:1053.4<br>M3:298.2 | M1: 353<br>M2: spherical<br>M3: 228 |
| Sensor Shape | | Toroid | Toroid | **Flat** |

## Discussion

In this work, we presented a 'first time right' design framework for unobscured and well-constraint four-reflection freeform imaging systems. In a first example, we demonstrated the effectiveness and speed of the proposed method for a 'conventional' four-mirror telescope with a wide rectangular FOV of 8.5º × 25.5º, which was inspired by reference literature[41]. In a step further, we reintroduced and proposed a novel class of multi-reflection imaging systems through the specific example of a two-freeform, three-mirror, four-reflection system. When compared with common three-mirror and three-reflection imaging systems[40], this multi-reflection approach shows unprecedented possibilities for both economic implementation and volume reduction. On top of the two examples, the introduced utmost flexibility to build metrics such as enclosed volume, obscuration, freeform PV departures, spot sizes, distortion or telecentricity, allow an excellent balance between manufacturability and image performance, which is very important for any feasible design. Furthermore, although the freeform surfaces are currently represented by XY polynomials, they can be readily converted to alternative freeform surface types such as Forbes polynomials[22] or Zernike polynomials[24] which has no impact on the achieved good performance.

Without loss of generality, our 'first time right' method can be applied to system designs with a flexible number of mirrors and/or lenses and further multi-reflection layouts. Its core strength lies in the rapid generation of initial design solutions with already significantly improved imaging performance while working with a largely reduced input parameter space. Note that one single 'first time right' calculation takes only dozens of milliseconds, it is very fast to automatically generate large quantities of qualified starting designs for screening, therefore it is perfectly suitable to be used as a global searching method. In general, a complete design takes typically less than five minutes (Intel i7-8850H CPU, 16GB RAM laptop) including an up to 500-loop optimization of the predefined positions within a certain range and evaluating the performance metrics. As a benchmark, another state-of-the-art automated design method for designing three-mirror freeform systems takes 2~3 hours[18]. Due to the fast speed, it is possible to investigate numerous starting geometries in a short time and evaluate them to find the optimal geometry for large FOV and/or fast mirror systems. As such, it is an ideal design method not only but especially for designers of all experience levels who want to rapidly explore new layouts and design solutions.

The presented 'first time right' design workflow has the full potential to provide a thorough and fast solution space search for a wide range of applications. Throughout the development of these two examples, we have evaluated various geometries and found that the initial geometry played a very critical role in achieving an optimal off-axis freeform imaging system. Future work will include applying advanced optimization algorithms to enable a further automated global search.